\documentclass[10pt,letterpaper]{article}
\usepackage{opex3}
\usepackage{multirow}
\usepackage{graphicx}
\newcommand{\expect}[1]{\left\langle #1 \right\rangle}

\begin{document}

\title{Yield enhancement in whispering gallery mode biosensors: microfluidics and optical forces}

\author{Kiran Khosla, Jon D. Swaim, Joachim Knittel and Warwick P. Bowen}
\address{Department of Physics, University of Queensland, St Lucia, QLD 4072 Australia}
\email{wbowen@physics.uq.edu.au} 

\begin{abstract*}

A microfluidic whispering gallery mode (WGM) biosensing system is proposed for enhanced delivery and detection of target molecules.  A microtoroid resonator coupled to a tapered optical fiber is immersed within a microfluidic channel, and supplied with target molecules at various flow rates.  We show through Monte Carlo simulations that the flow characteristics and resonantly enhanced optical forces of the sensor substantially improve both the sensing time and yield.  When compared to a diffusion-limited sensing modality, the average time required to detect a single molecule is reduced from more than 100 minutes to less than 10 seconds, and the overall yield of the device is enhanced from less than 5\% to a maximum of 70.6\% for femtomolar concentrations of analyte.  

\end{abstract*}

\vspace{7mm}

\hspace{8mm} \copyright 2010 Optical Society of America




\section{Introduction} 

Whispering gallery mode (WGM) resonators are optical sensors that exhibit unprecedented sensitivity due to their small mode volume and ultra-high Q factor (Q$\sim 10^8$)~\cite{Vahala:03, Kippenberg:04, Spillane:04}. In the life sciences, they have been shown to be excellent candidates for biological sensing~\cite{Joachim, Arnold:Protein_Shift03, Vollmer:02}, and have recently demonstrated single molecule sensitivity by detecting individual proteins~\cite{Armani:07} and virus molecules~\cite{Vollmer:08} without the use of labels.  In a WGM biosensor, light is evanescently coupled to a water-immersed resonator via a tapered optical fiber.  The resonant frequency of the light circumnavigating the resonator is sensitive to small dielectric pertubations in its environment, such as the adsorption of proteins on the resonator's surface.  Detection of single biomolecules is achieved by functionalizing the surface of WGM sensors with anti-bodies or DNA specific to the target molecule, and then monitoring the resonance frequency spectrum of the sensor.  

The performance of a biosensor, however, is largely dependent on the transport and binding kinetics of target molecules at the sensor's surface.  Generally, the transport process at the surface is exclusively diffusive, and therefore the measurement of sub-picomolar concentrations without directed transport can require time scales of hours to days~\cite{Squires}.  Introducing a simple convective flow over the surface does not appreciably improve the yield of the device, as diffusion tends to dominate at the sensor surface and convection dominates further away.  Since 
it is often desirable to analyze small volumes or extremely dilute concentrations of analyte~\cite{Squires}, an efficient method of delivering target molecules to the sensor's surface, such as a microfluidic system, is needed to ensure an acceptable yield in a practical biosensor.

In this paper we propose to integrate a microtoroid WGM sensor into a microfluidic system that transports molecules directly towards the most sensitve area of the sensor, and then captures them by the gradient forces of the resonantly enhanced electromagnetic field.  We show that these processes combined can overcome the diffusion-limited transport mechanism and reduce the time required to detect femtomolar concentrations of molecules from more than 100 minutes in a diffusion-limited sensor to less than 10 seconds.  As expected, with increasing convection a compromise is observed between improved sensing time and reduced yield due to the shorter time spent by molecules near the sensor surface.  However, we show that this compromise between yield and fast detection time can be overcome by designing a microfluidic channel geometry that induces vortices for large flow rates, acting to focus the molecules onto the sensing region.  Significant increases in the yield are then observed over a wide parameter regime, with a yield as high as 40\% achieved with a 7 second average sensing time.        

\section{Theory}

In our proposed sensing system, a molecule of radius $a$ moves toward the WGM sensor with a velocity relative to the fluid flow $\dot{x}(t)$ and experiences a viscous force proportional to its velocity by $\beta = 6\pi\eta a$, where $\eta$ is the viscosity of the solution.  It also experiences a Brownian force $\sqrt2k_bT \gamma(t)$, where $k_b T = $ 4.11 x 10$^{-21}$ J at room temperature, and an attractive optical force

\begin{equation}
F(r) = \frac{1}{2}\alpha_{ex}\nabla |\mathbf{E}(\mathbf{r})|^2
\label{eq:force_grad}
\end{equation}

\noindent from the gradient of the evanescently decaying electromagnetic field $\mathbf{E}(\mathbf{r})$ within the WGM sensor, where $\alpha_{ex} $ is the excess polarizability of the molecule.  Thus, the molecule follows a trajectory given by 

\begin{equation}
\ddot{x}(t) =  -\frac{\beta}{m}\dot{x}(t)-\frac{F(x)}{m} + \frac{\sqrt{2\beta k_B T}}{m}\gamma(t)
\label{eq:motion}
\end{equation}

\noindent where $m$ is the mass of the molecule.  When the molecule reaches the sensor, the optical gradient force $F$ pulls it towards a region of high field intensity.  If the molecule adsorbs onto the surface, the evanescent field polarizes the molecule, inducing an oscillating dipole moment $\delta p=\alpha_{ex}\mathbf{E}(\mathbf{r})$ in the molecule, resulting in a shift of the WGM resonance frequency $\omega_0$ given by ~\cite{Arnold:Protein_Shift03}:
\begin{equation}
\delta \omega = \frac{-\alpha_{ex}|\mathbf{E}(\mathbf{r})|^2}{2V_m|\mathbf{E}_{max}|^2} \omega_0
\label{eq:freq_shift}
\end{equation}
\noindent where $V_m$ is the volume of the optical mode and $\mathbf{E}_{max}$ is the maximum intensity of $\mathbf{E}(\mathbf{r})$.  The minimum frequency shift observable with a given WGM sensor will depend on a range experimental parameters, including most critically the laser linewidth and power, the WGM $Q$ factor, and whether the laser is frequency scanned or locked on resonance.  Since detection of sub-linewidth resonance shifts is possible when frequency locking and stabilization methods are employed~\cite{Black:PHD}, in the article we select a minimum frequency shift of 1.75 kHz for a detectable binding event, roughly 1/10 of the linewidth of a tunable diode laser recently developed by D$\ddot{\mathrm{o}}$ringshoff \emph{et al.}~\cite{Doringshoff:07}.

The transport characteristics of the biosensor for a given flow field are determined by numerically solving Eq.~(\ref{eq:motion}) to find the expected position of the molecule $\expect{x(t)}$.  If the time interval $dt$ is small enough that $F(x)$ is approximately constant over the interval and the Brownian term $\expect{\gamma(t)} = 0$, a finite difference method provides the next expected position $\expect{x_{n+1}}$ and velocity $\expect{\dot{x}_{n+1}}$ of the molecule

\begin{eqnarray}
\expect{x_{n+1}} &=& \frac{m}{\beta}\left[\left(\frac{F}{\beta} - \dot{x}_n \right)\exp{\frac{-\beta dt}{m}}-\frac{F}{\beta}+\dot{x}_n \right] + \frac{F dt}{\beta} + x_n  \label{eq:OdeSolX}\\
\expect{\dot{x}_{n+1}}&=& \left(\frac{F}{\beta} - \expect{\dot{x}_n}\right )\exp{\frac{-\beta dt}{m}} + \frac{F}{\beta}
\label{eq:OdeSolV}
\end{eqnarray}

\noindent Brownian motion is taken into account by adding Gaussian noise terms with mean zero to the position and velocity~\cite{Uhlenberk:30}.  These terms have variances of $\sigma^2 = \frac{2k_b T}{\beta}\left[\delta t - \frac{m}{\beta}\left(1-\exp{-\frac{\beta}{m}\delta t}\right)\right]$ and $\sigma_v^2 = \frac{k_bT}{m}$, respectively.  When a noise term is added to the position, there is some correlated noise in the velocity given by $\delta \dot{x} = \delta \overline{\dot{x}} + A\delta{x}$, where $\delta\overline{\dot{x}}$ is the truly random part of the velocity kick and $A$ is the correlation coefficient of the velocity kick in the direction of $\delta x$.  By multiplying the noise in the velocity by $\delta x$ and taking the expectation value, we find $\expect{\delta x\delta\dot{x}} = \expect{\delta x \delta\overline{\dot{x}}} + A\expect{\delta x^2}$.  This ordinary differential equation can then be solved to the find the correlation coefficient $A_{n+1}$ at any step $n+1$ given an initial correlation constant $A_n$.
\begin{equation}
A_{n+1} = \frac{\left[\left(A_n + \frac{k_b T}{\alpha}   \right)  \exp\left(\frac{-\alpha dt}{m} \right) +  \frac{k_b T}{\alpha}\right]} {\frac{2k_b T}{\alpha}\left\{ t - \frac{m}{\alpha}\left[1-\exp\left(\frac{-\alpha dt}{m}  \right)\right]\right\}}
\label{eq:A}
\end{equation}
\noindent Since the variance of the truly random velocity kick $\expect{\overline{\dot{x}}}$ is known, we can then add a Brownian term to the new velocity $\expect{\dot{x}_{n+1}}$ given by $A_n\expect{\delta x} + N(0, \expect{\overline{\dot{x}}})$.   

\section{Modeling}

\begin{figure}[h!]
	\begin{center}
	\includegraphics[scale=0.4]{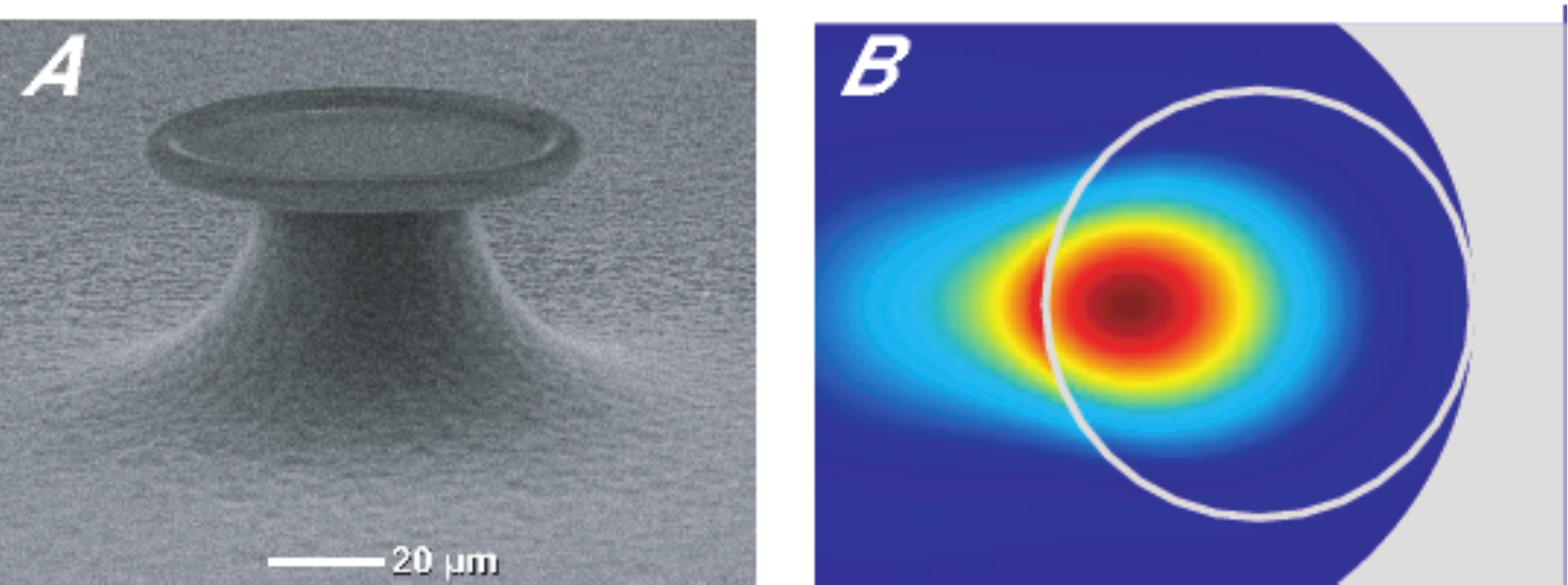}
	\caption{(A) SEM image of a microtoroid.  Scale bar is 20 $\mu$m.  (B)  Vertical cross-section of the finite element model of the electric field within a a 20 $\mu$m diameter toroid immersed in water.  The circle indicates the boundary of the microtoroid, and the grey area represents the supporting silicon pedestals.}
	\label{fig:toroid}%
	\end{center}
\end{figure}

Silica microtoroids were chosen as the WGM resonators that constitute the basis of our biosensing system, and are fabricated with a combination of photolithography, dry etching and a selective reflow process~\cite{Vahala:03}.  The optical mode volume is much smaller than that of its microsphere counterpart, and therefore exhibits a higher sensitivity to frequency shifts resulting from molecular binding events (Eq.~\ref{eq:freq_shift}) as well as a larger gradient force (Eq.~\ref{eq:force_grad}). A typical microtoroid is shown in Figure~\ref{fig:toroid}A.  We use a finite element model (COMSOL Multiphysics\textregistered \hspace{1pt} 3.4) to numerically solve for the fundamental transverse magnetic (TM) mode in a water-immersed microtoroid with major and minor diameters of $D$=20 $\mu$m and $d$=6 $\mu$m, respectively.  Figure~\ref{fig:toroid}B shows the extent of the evanescent field into the aqueous environment, and illustrates the basis for biological sensing using WGM microcavities.  To approximate the mode volume of a microtoroid, Spillane~\emph{et al.}~\cite{Spillane:04} derived a simple equation to express the mode volume in terms of its "sphere-like" and "step-index fiber-like" components

\begin{equation}
V_m(d,D) \simeq \left[(\pi DA_{m}^{fiber}(d))^{-3}+\left((\frac{d}{D})^{1/4}V_m^{sphere}(D)\right)^{-3}\right]^{-1/3}
\label{eq:mode_volume}
\end{equation}

\noindent where $A_{m}^{fiber}$ is the mode area of a step-index fiber with diameter $d$ and $V_m^{sphere}$ is the mode volume of sphere with diameter $D$.  Using similiar values as in Ref~\cite{Spillane:04}, this expression approximates the mode volume of a microtoroid with an error less than 4\% and, in our case, gives a value of $V_m$=98.4 $\mu$m$^3$.  From Eq.~(\ref{eq:freq_shift}), the maximum frequency shift achievable for a single adsorbed molecule can then be estimated.   We consider the protein Bovine Serum Albumin (BSA), with a polarizability $\alpha_{ex}=3.85 \times 10^{-27}$ m$^3$.  Due to the electromagnetic boundary conditions, a discontinuity is present in the electric field at the microtoroid surface, as shown in Figure~\ref{fig:toroid}B, enhancing the molecule-resonator interaction.  As a result, we find from the numerical solutions for the electric field given by COMSOL that $\textbf{E}(\textbf{r})$ is maximum at the equatorial surface of the microtoroid, giving a ratio $|\textbf{E}(r)|^2/|\textbf{E}_{max}|^2$ of one.  The maximum frequency shift is then predicted to be 6.8 kHz, larger than our minimum detectable shift defined previously.

\begin{center}
	\begin{figure}[ht!]
		\begin{center}
			\includegraphics[scale=0.3]{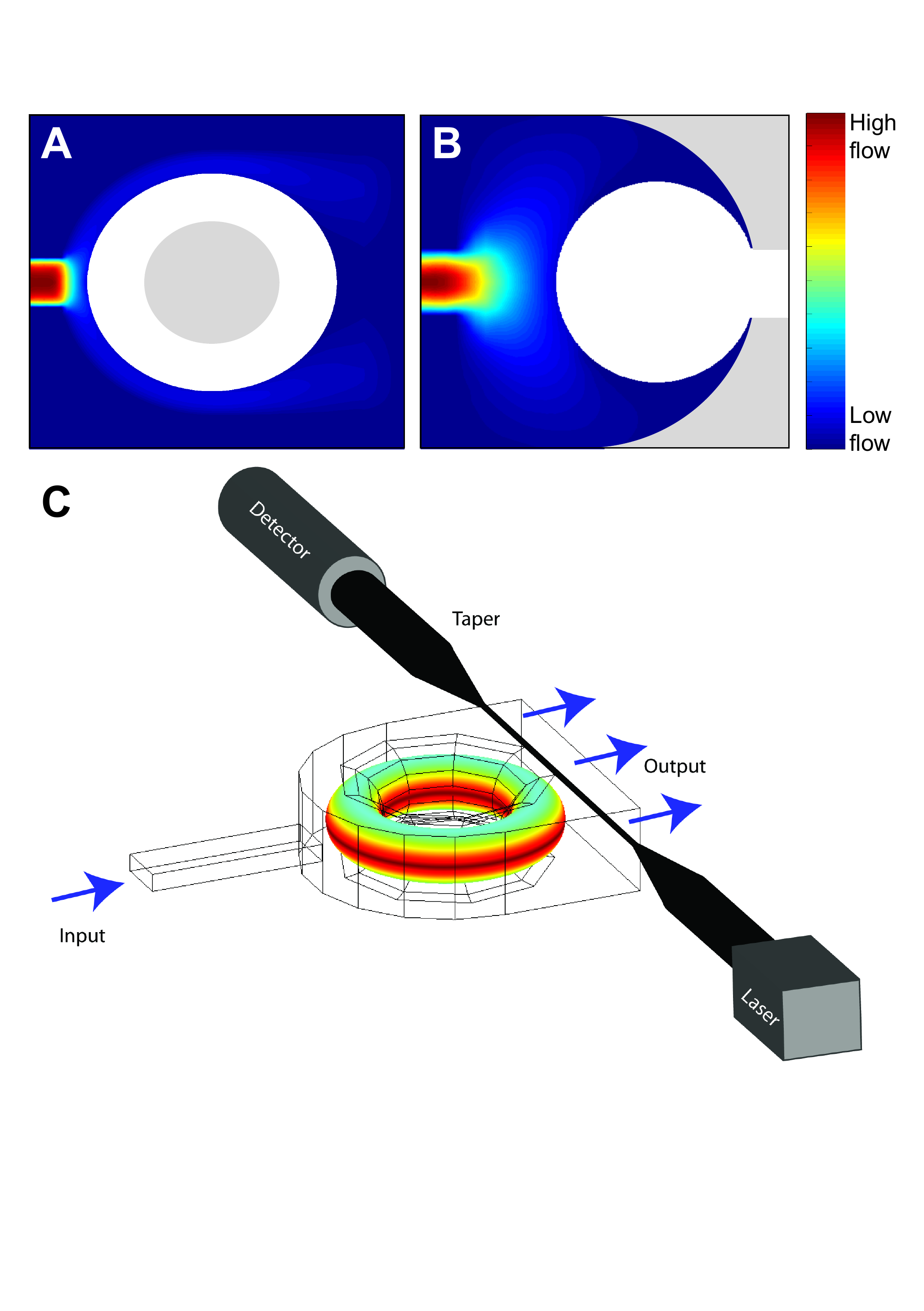}
			\caption{(A) Vertical and (B) horizontal cross-sections of the flow field for an input pressure of 110.2 kPa.  White and grey areas indicate the interior of the microtoroid and its supporting pedestal, respectively.  (C) A microtoroid coupled to a tapered optical fiber and incorporated into the microfluidic system. The intensity map around the microtoroid indicates the intensity of the TM mode at the surface.}
			\label{fig:setup}%
		\end{center}
	\end{figure}
\end{center}

As shown in Figure~\ref{fig:setup}C, the microfluidic system considered here consists of a 6 $\mu$m wide, 2 $\mu$m tall input channel that infuses a 11.6 $\mu$m tall chamber which contains the silica microtoroid.  The input channel has a length of 4 $\mu$m.  Near the input channel, the chamber is semi-circular with a radius of 16 $\mu$m, and thus encircles half the microtoroid.  At a distance of 16 $\mu$m from the input channel, the chamber remains 32 $\mu$m wide and functions as a large rectangular ouput channel with a length of 16 $\mu$m.  The total volume of the microfluidic system is roughly 6.75 pL.  COMSOL is used to characterize the fluid flow field within the chamber, with the geometry defined by a mesh of 5811 points (26305 elements and 3940 boundary elements) and the entrance pressure at the input channel defined relative to atmospheric pressure as a boundary setting.  The flow field around the microtoroid is found by solving the incompressible Navier-Stokes equations.  Typical horizontal and vertical cross-sections of the flow are shown in Figures~\ref{fig:setup}A and B, respectively.  The flow is quadratic within the input channel, and then spreads out into the chamber, giving a low flow field at the surface of the microtoroid and ultimately atmospheric pressure at the outlet.

Although the fabrication details of the microfluidic system are not discussed in this paper, a flip-chip-based design  is proposed as a feasible method to enclose the microtoroid, with the microfluidic chamber defined by patterned channels in two identical silicon chips.  We envisage removing the microtoroid from the top chip via wet etching, leaving the silicon pedestal which can then be placed down upon the top of the microtoroid on the bottom chip (Figure~\ref{fig:setup}C).  As a result, the final device consists of a microtoroid sandwiched by two silicon pedestals (grey areas in Figure~\ref{fig:setup}B).

To analyze the transport characteristics of our biosensor, we perform Monte Carlo simulations of BSA proteins passing through the microfluidic system, governed by the equation of motion in Eq.~(\ref{eq:motion}).  A total of 10,000 molecules with initial velocities $\dot{x}_{0}$ taken to be their mean velocities relative to the fluid flow $\dot{x}(t)$ are allowed to pass through the system until the molecules bind to the microtoroid or flow past it.  In the simulation we do not explicity consider the density of binding sites ($\sim$5 x 10$^{14}$ cm$^{-2}$)~\cite{Zhuravlev} on the surface of silica.  Rather, we consider a molecule bound to the microtoroid when it passes through the mesh boundary defining the interface of the chamber and microtoroid, after which it is not allowed to de-bind from the surface.  This is a reasonable approximation in the case of low molecular concentrations, where the sensor binding sites are far from saturated.  Indeed, our simulations are well within this regime when only 10,000 molecules are present.  Thus, the kinetics of our binding events are expected to be linear, rather than following the typical behavior of first-order Langmuir kinetics~\cite{Squires}.  A time step of 2 $\mu$s is used in the finite difference method to track the molecules' positions in time, and the frequency shifts due to the binding events are calculated from Eq.~(\ref{eq:freq_shift}).   To investigate the effect of convection on binding kinetics we perform a series of simulations in which the pressure difference between the input channel and the outlet ranges from 0 to 190 kPa, where a difference of 10 kPa roughly corresponds to a flow rate of 0.4 $\mu$L/min. 

For comparison with a diffusion-limited sensing modality, we perform additional simulations for a second geometry in which a microtoroid is placed in a micro-droplet of solution with a volume V$_{d}$ of 1 $\mu$L.  Since 10,000 molecules suspended in a 1 $\mu$L solution give a concentration of about 16 fM, our micro-droplet is representative of a typical miniaturized sensor with respect to volume and concentration~\cite{Squires}.  The volume of the micro-droplet is fixed so that, for comparison, the time of each diffusive simulation given by V$_{d}$/q$_f$ is the same as the microfluidic case, where q$_f$ is the flow rate of the corresponding microfluidic simulation.  Thus, we can compare the efficacy of both systems in detecting single molecules for a given sensing time.  For both geometries, the optical power carried by the tapered optical fiber is varied from 0 mW to 10 mW, and the operating wavelength chosen for the study is 852.4 nm.  

\section{Results}

\begin{figure}[ht!]
		\begin{center}
			\includegraphics[scale=0.4]{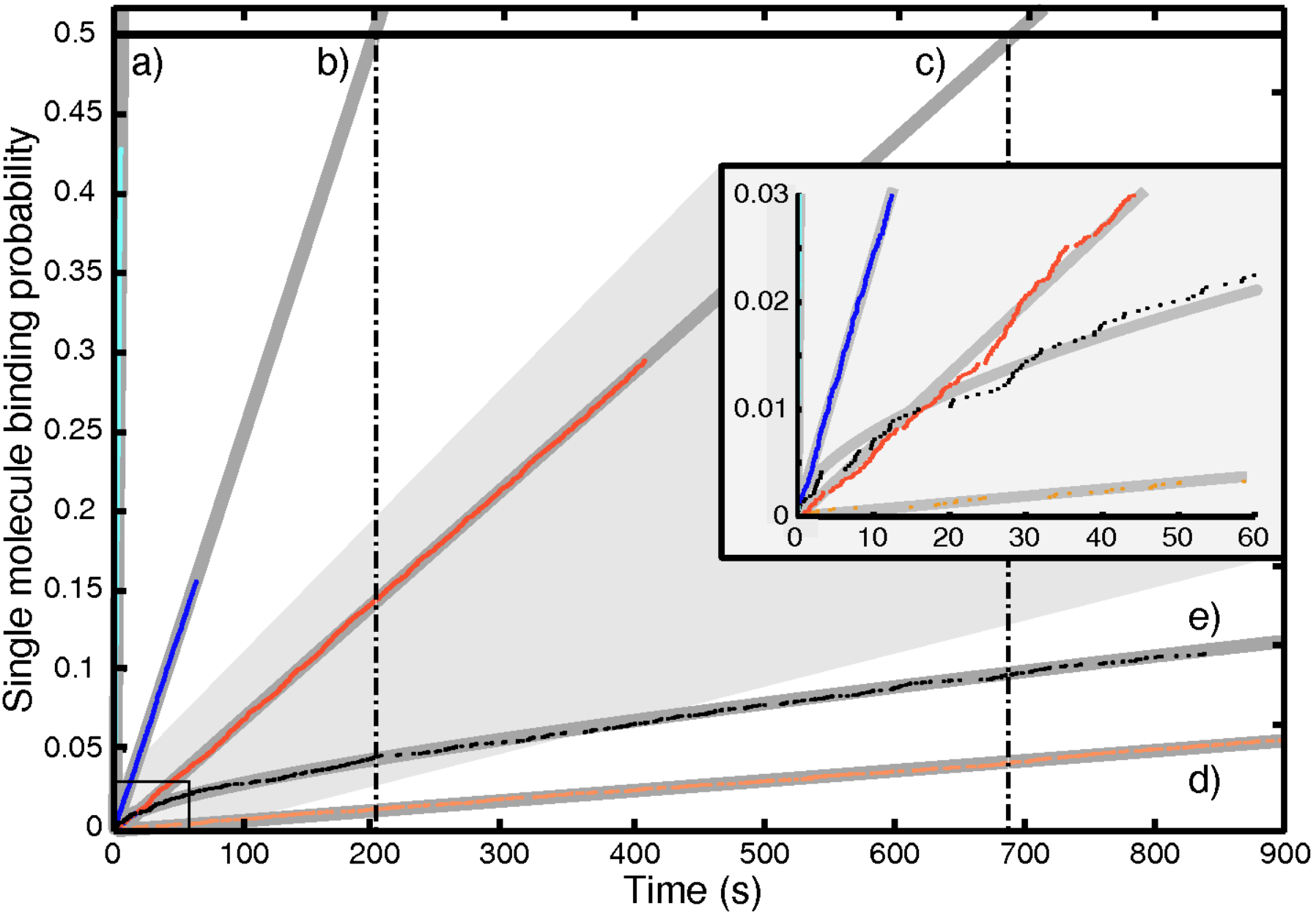}%
			\caption{Single molecule binding probability for the microfluidic and diffusion sensing systems.  The input pressures are (a) 273.2 kPa, (b) 113.2 kPa, (c) 103.2 kPa, (d) 101.4 kPa, and (e) corresponds to the response of the diffusion system.  The inset shows the transient response of the sensor at short time scales.}%
			\label{fig:kinetics}%
		\end{center}
\end{figure}

In this paper we characterize the efficacy of our WGM bioensor with three properties: the yield, or percentage of molecules detected; the efficiency, defined as the probability of detection once a molecule reaches the sensor; and the detection time, or the average time needed for a molecule to be detected.  In Figure~\ref{fig:kinetics}, we show the  number of binding events as a function of time, normalized to the total number of molecules for our microfluidic and diffusion sensing systems at a fixed laser power of 1 mW.  Normalized in this way, the vertical axis corresponds to the probability of an initially randomly placed molecule successfully binding to the sensor.  The average single molecule detection time is found by extrapolating this data to find the time at which the binding probability is 50\%.  

In the case of diffusion (Figure~\ref{fig:kinetics}e), a thin depletion zone develops around the surface of the sensor and grows as $\sqrt t$ when anisotropy in diffusion is neglected~\cite{Squires}.  Initially, the detection at the sensor surface also grows as $\sqrt t$ until steady-state accumulation is reached and the detection becomes linear with time~\cite{Whitman}.  By contrast, in the microfluidic system, the sensor response is linear with time even for small times and rapidly increases with increasing flow rates.  Table~\ref{efficiency1} compares the efficiency and average sensing times of the microfluidic and diffusion systems for a laser power of 1 mW.  Importantly, the average time required for one molecule to bind with a 50\% probability can be reduced by three orders of magnitude simply by varying the pressure difference between the inlet and the outlet of the microfluidic chamber.  We note that the associated flow rates are easily attainable in typical experimental conditions using a syringe pump, and that the introduction of a modest flow can enable detection of femtomolar concentrations on a sub-100s time scale.  Furthermore, over 90\% efficiency is maintained for the microfluidic system regardless of input pressure, whereas a much lower efficiency  (36\%) is observed in the diffusion system.

\begin{table}[]
\caption{Comparison of the performance of microfludic and diffusion systems as a function of input pressure.  The incident optical power is 1 mW.}
\begin{center}
\begin{tabular*}{0.6\textwidth}{@{\extracolsep{\fill}}ccc}
\hline
\multirow{2}{*}{} Pressure difference & Efficiency & Average \\ (kPa) & & sensing time (s) \\
\hline
0* & 0.90 & 7540 \\
1.8 & 0.92 & 671  \\
11.8 & 0.94 & 200  \\
51.8 & 0.94 & 48  \\
81.8 & 0.94 & 18  \\
171.8 & 0.92 & 7 \\
\hline
Diffusion Only & 0.36 & 6032 \\
\hline
\multicolumn{3}{l}{* Atmospheric pressure; transport is exlusively diffusive.}
\end{tabular*}
\end{center}
\label{efficiency1}%
\end{table}

\begin{figure}[hb!]
		\begin{center}
			\includegraphics[scale=0.65]{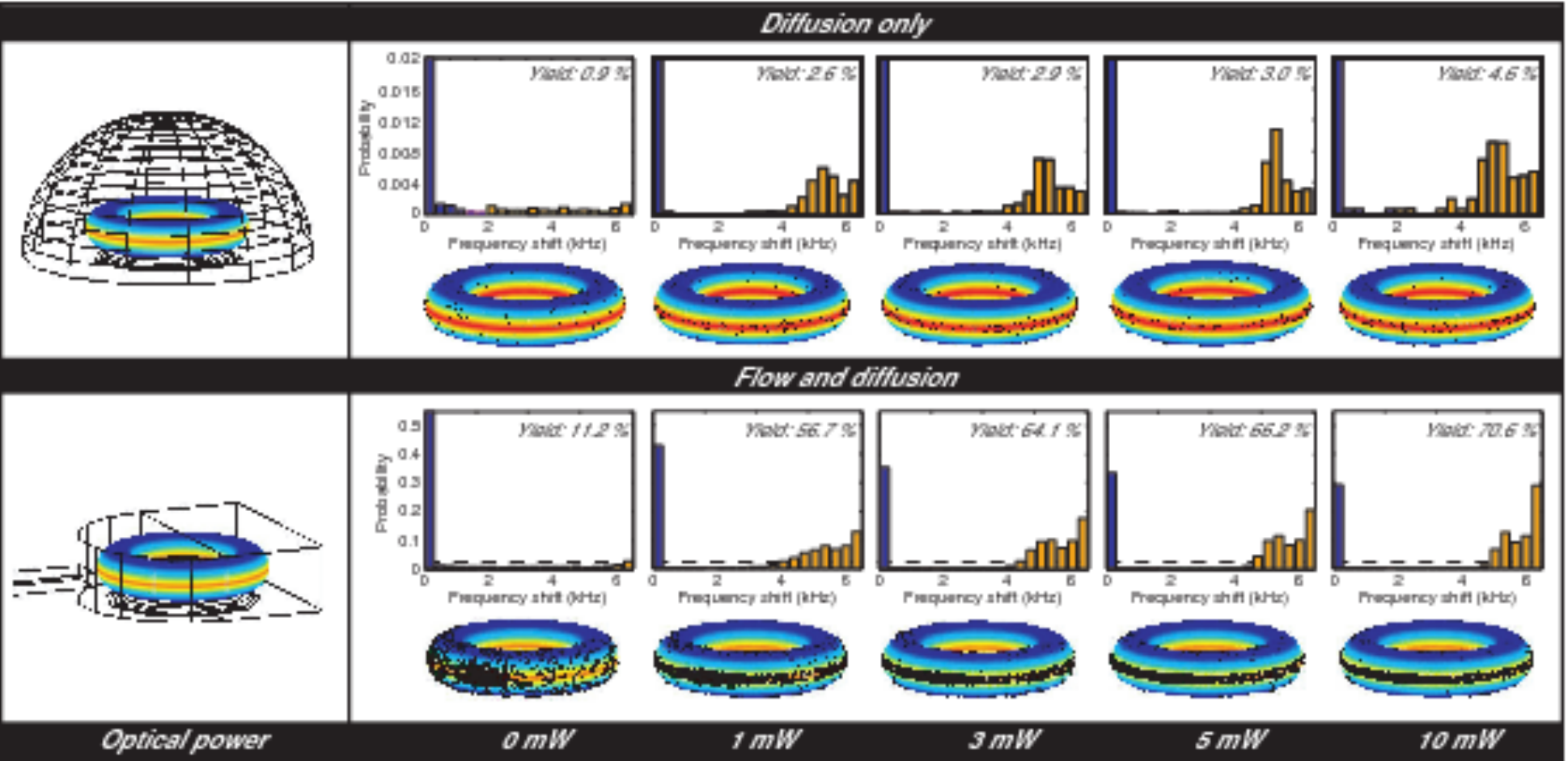}%
			\caption{Comparison of the percentage of molecules detected in the microfluidic and diffusion systems within 10 seconds.  Figures at the top and bottom show the distributions of frequency shifts and calculated yields as a function of power for the diffusive and microfluidic systems, respectively.  The pressure difference in the microfluidic system is 0.8 kPa, and the bin size of the histograms is 2 kHz.}%
			\label{fig:histograms}%
		\end{center}
\end{figure}

An important feature of any WGM-based sensor is its ability to attract molecules via the gradient of the evanescent optical field.  In Table~\ref{efficiency2}, we show that the optical field can significantly reduce the average sensing time of the microfluidic system, and that over 90\% efficiency is maintained for optical power of 1 mW or more.  In Figure~\ref{fig:histograms}, we show the effect of increasing the input laser power on the overall yield of the sensor for a given flow rate and sensing time.  In the microfluidic system, the presence of 1 mW of incident optical power increases the yield from 11.2\% (0 mW) to 56.7\%, illustrating the sheer practicality of employing optical cavities as biosensors. It is important to note that most experiments to date~\cite{Armani:07, Vollmer:08, Arnold:Protein_Shift03} have operated close to 0 mW, since detection was achieved via repeated scanning of the laser over the resonance frequency as opposed to locking to it, resulting in a very low average power within the resonator.  Increasing the circulating optical power results in a larger evanescent wavefront, attracting molecules further away from the sensor throughout the duration of the simulation.  The yield therefore increases as a function of power, and reaches a maximum of 70.6\% for 10 mW of incident power.  By comparison, over the same sensing time scale, less than 5\% yield is observed when transport is diffusion-limited, regardless of optical power (top of Figure~\ref{fig:histograms}).

\begin{table}[t!]
\caption{Performance of the microfluidic system as a function of optical power for a constant pressure difference of 0.8 kPa.  }
\begin{center}

\begin{tabular*}{0.6\textwidth}{@{\extracolsep{\fill}}ccc}
\hline
\multirow{2}{*}{} Power & Efficiency & Average \\ (mW) & & sensing time (s) \\
\hline
0 & 0.38 & 1421 \\
1 & 0.92 & 687 \\
3 & 0.95 & 621 \\
5 & 0.96 & 607 \\
10 & 0.97 & 581 \\
\hline
\end{tabular*}
\end{center}
\label{efficiency2}%
\end{table}

A corollary of the optical force is the redistribution of frequency shifts seen in the histograms of Figure~\ref{fig:histograms}.  A stronger optical field attracts molecules faster, giving them less time to diffuse away in the azimuthal direction.  Consequently, molecules are more likely to bind to areas of the sensor with high field strength.  In our simulation, this translates into larger frequency shifts from Eq.~(\ref{eq:freq_shift}), and an increase in the number of counts in the maximum frequency shift bin (6.2 kHz).  The toroidal images in Figure~\ref{fig:histograms} show the final locations of the bound molecules, and depict the equatorial localization of bound molecules observed in the simulations.

We find that the microfluidic system always produces a higher yield than its diffusive counterpart.  This can be attributed to the shape of the flow profile after leaving the channel, which acts to force molecules onto the sensing area.  If the flow rate $q_f$ is sufficiently large in the microfluidic system, it might be expected that some molecules flow past the sensor faster than they are able to diffuse towards it, and therefore fail to bind, leading to a reduction in yield.  Figure ~\ref{fig:pressure} displays the calculated yields for our biosensor as a function of input pressure with 0 mW (yellow squares) and 1 mW (blue circles) of optical power.  As expected, the yield sharply declines with increasing pressure, since more molecules flow past the microtoroid undetected due to the increasing dominance of convection over diffusion.  As the pressure is increased above 140 kPa, however, vortices develop in the corners of the chamber and input wall, inducing a convective flow that directs molecules back towards the front wall of the sensor, as illustrated in Figure~\ref{fig:pressure}B and C.  Fortuitously the re-directed flow increases the yield of the sensor, which steadily increases with pressure until a threshold around 275 kPa is reached, after which the flow is again less directive and flows past the microtoroid.  Thus, careful choice of the microfluidic channel design allows a very short measurement time to be achieved at high pressure without compromising the yield of the device.  At 273.2 kPa, a detection yield of 40\% is achieved with a sensing time of 7 seconds (Table~\ref{efficiency1}) for only 1 mW of input power.  As can be seen in Figure~\ref{fig:pressure}, the optical force plays less of a role and ceases to increase the yield as fewer stray molecules flow past the microtoroid, with the ratios of yields for 1 mW and 0 mW reaching unity at pressures greater than $\sim$180 kPa.

Although the greatest yield of 70.6\% is achieved for low convection and a long sensing time, the yield associated with high convective flows can be increased by recirculating the flow back into the microfluidic system, so that undetected molecules may be subsequently captured.  Thus, one would expect an enhancement in the yield increasing with the number of recirculations $j$, allowing for a very high yield $Y^j$ to be achieved with a low measurement time:  

 \begin{center}
	\begin{figure}[ht!]
		\begin{center}
			\includegraphics[scale=0.75]{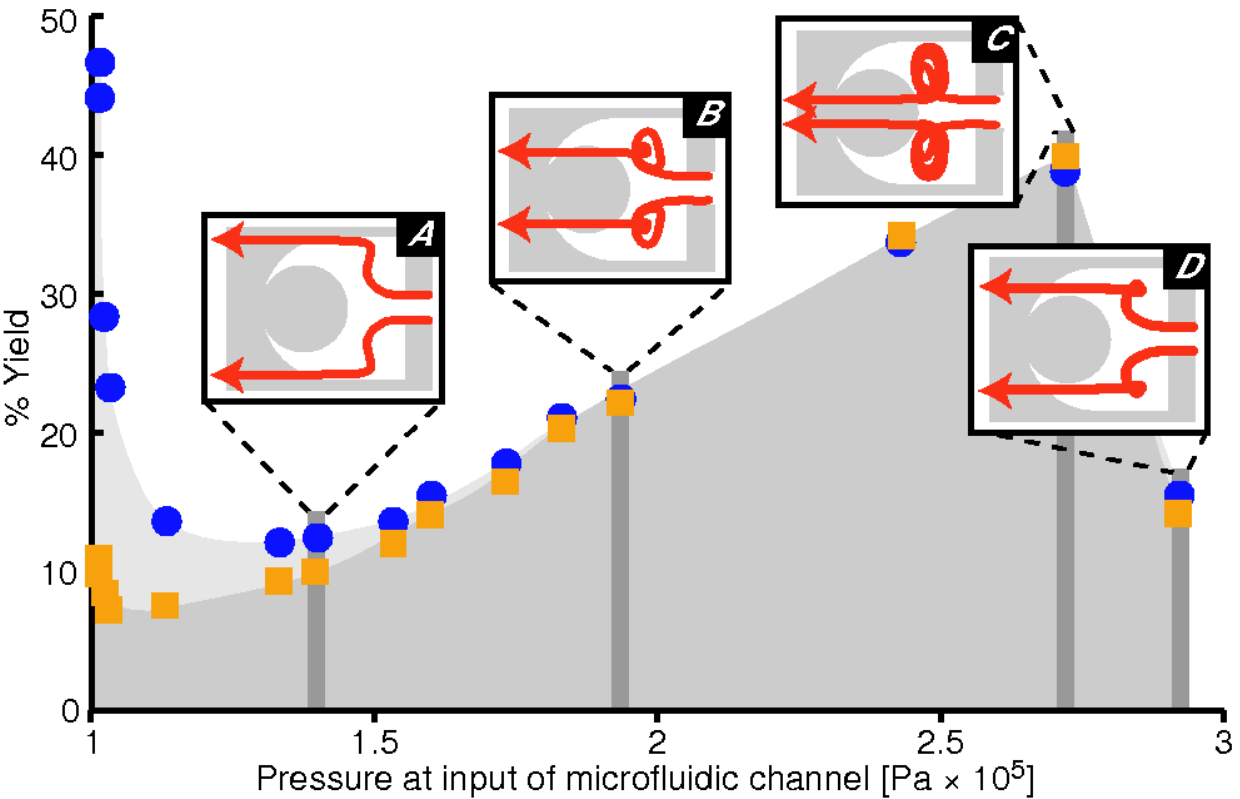}%
			\caption{The effect of pressure on the yield of the microfluidic system. Insets A-D depict the trajectory of the molecules in the microfluidic chamber, illustrating the rough characteristics of the flow field within the chamber for the range of flow rates used in this study.  The incident optical power is 0 mW (yellow squares) or 1 mW (blue circles).}%
			\label{fig:pressure}%
		\end{center}
	\end{figure}
\end{center}   

\begin{equation}
Y^j=\frac{N_{detected}}{N_{in}}=\eta \left[1-\left(1-\frac{Y}{\eta} \right)^j \right]
\label{eq:yield_j}
\end{equation}

\noindent where $\eta$ is the efficiency of the sensor, and $N_{detected}$ and $N_{in}$ are the number of detected and incident target molecules, respectively.  Note that for an infinite number of recirculations, the yield approaches the efficiency as expected, and that $Y^j$ is simply the inital yield for one recirculation ($j$=1).  For high convective flows such as the case when the input pressure is 273.2 kPa, we calculate that the initial yield of 40\% can be increased to more than 75\% with only three recirculations (and in only 21 seconds).  Thus, our microfluidic system is capable of achieving very high yields for very sensible time scales.  Furthermore, we note that, even without recirculation, just 1 mW of optical power can increase the yield of the microfluidic device to more than 50\% for slow flow rates, more than 20$\times$ greater than the yield from the diffusion-limited case.

\section{Conclusion}
We have proposed a WGM biosensing sytem that greatly improves the detection of molecules in solution.  The chief advantage in using WGM-based resonators as advanced biosensors is that single molecules can be detected without the use of labels and with very low input power.  Generally, the utility of WGM sensors would be limited by the confinement of light to the small volume of the resonator.  However, we have shown that incorporating a microtoroid WGM sensor into a microfluidic system allows for controlled manipulation of molecules and its surrounding fluid, substantially improving the probability of detection over a reasonable time frame.  Single molecules can be detected with an average measurement time as low as 7 seconds, and the overall yield of the device can be enhanced from less than 5\% in the case where diffusion dominates to as high as 70.6\% where the interplay of convection and optical forces becomes dominant.  Particles or molecules as small as a single BSA protein (roughly 6 nm in radius) can be detected, and the analysis can be completed for femtomolar concentrations.  These results demonstrate that the combination of microfluidics and WGM sensing provides a highly tunable system, with both yield for a given concentration and sensing time optimizable by varying the input power and flow characteristics of the microfluidic system.  

\section{Acknowledgments}
This research was funded by the Australian Research Council Grant No. DP0987146.

\end{document}